\documentclass[preprint,5p,twocoloumn]{elsarticle}
\journal{Int. J. Mod. Phys. B}
\usepackage{amsfonts,physics}
\usepackage[utf8]{inputenc}
\usepackage{hyperref}
\hypersetup{colorlinks=true,urlcolor=red, citecolor=blue,linkcolor=blue}
\allowdisplaybreaks
\usepackage{amsmath}
\allowdisplaybreaks
\usepackage{amsfonts}
\usepackage{amssymb}
\usepackage{graphicx}
\usepackage{float}
\begin{document}
	\title{Nonclassicality in a dispersive atom-cavity field interaction in presence of an external driving field}
	\author{Naveen Kumar}
	\ead{naveen74418@gmail.com}
	\author{Arpita Chatterjee\corref{cor1}}
	\ead{arpita.sps@gmail.com}
	\cortext[cor1]{Corresponding author}
	\date{\today}
	\address{Department of Mathematics, J. C. Bose University of Science and Technology,\\ YMCA, Faridabad 121006, India}
	\begin{abstract}
		We investigate nonclassical properties of a state generated by the interaction of a three-level atom with a quantized cavity field and an external classical driving field. In this study, the fields being degenerate in frequency, are highly detuned from the atom. The atom interacts with the quantized field in a dispersive manner. The experimental set-up involves a three-level atom passing through a cavity and interacting dispersively with the cavity field mode. Simultaneously, the atom interacts with an external classical field that is in resonance with the cavity field. The three-level atom can enter the cavity in one of the bare states $\ket{e}$, $\ket{f}$ or $\ket{g}$ or in a superposition of two of these states. In this paper, we consider superposition of $\ket{e}$ and $\ket{f}$. In our analysis, we focus on the statistical properties of the cavity field after interacting with the atom. The state vector $|\psi(t)\rangle$ describes the entire atom-field system but we analyze the properties of the cavity field independently neglecting the atomic component of the system. For this the atom part is traced out from $|\psi(t)\rangle$ to acquire the cavity field state only, denoted by $\ket{\psi_{ f}(t)}$. We evaluate different nonclassical measures including photon number distribution, Mandel's $Q_M$ parameter, squeezing properties $S_x$ and $S_p$, Wigner distribution, $Q_f$ function, second-order correlation function $g^2(0)$ etc. for the obtained cavity field state.

	\end{abstract}
	\maketitle
	\section{Introduction}
The well-known Jaynes-Cummings model (JCM) \cite{jcm} is a theoretical framework that describes the interaction between a two-level atom and a single-mode quantized cavity electromagnetic field \cite{qed} with the rotating-wave approximation. The JCM is considered to be the most fundamental model for studying the interaction between matter and field in the field of quantum optics. Moreover, this model has nonperturbative solutions that are exactly integrable. A large number of multi-level and multi-mode extensions of the original JCM have been studied over the years \cite{swore}. The driven Jaynes-Cummings model while the
cavity and the external driving field are close to or on resonance with the atom, has been studied by several authors. For example, Alsing et. al. \cite{alsing} studied the Stark splitting in the
quasienergies of the dressed states resulting from the presence of the driving field subject to the condition  that both driving and cavity fields are
resonant with the atom. Jyotsna and Agarwal \cite{jyostna} studied the
effect of the external field on the Rabi oscillations in a situation
where the cavity field is resonant with the atom and where
the external field is both resonant and non-resonant. Dutra et. al. \cite{dutra} studied a similar model but
where the external field was taken to be quantized. Chough
and Carmichael \cite{chough} have studied the JCM with an external resonant driving field and have shown that the collapses and
revivals of the mean photon number occur over a much
longer time scale than the revival time of the Rabi oscillations for the atomic inversion. Joshi \cite{joshi}, in a similar way,
studied the driven two-photon JCM when Nha et. al. \cite{nha} studied the preparation of a temporally stable single-photon state in an atom-cavity field system with a driving classical field. To the best of our knowledge, the dispersive interaction with an external driving field has not considered substantially despite that it is a logical extension of the previous work in this field. 

In the context of original JCM, if the interaction occurs in limit of a
	large (but not too large) detuning between the cavity field
	and the relevant atomic transition frequency, then it is called dispersive which is proved to be of great importance in various proposals and experiments for producing superposition of macroscopically (or at least mesoscopically) distinguishable quantum
	states, the so-called Schr\"{o}dinger cat states in cavity QED system. Dispersive atom-field interactions are utilized in quantum sensors, such as atomic magnetometers and atomic clocks. These interactions enhance the precision of measurements with reference to quantum metrology. They are used to design and develop quantum-enhanced measurement devices with applications in navigation, geophysics, and precision spectroscopy. Dispersive interactions can be utilized to create quantum amplifiers, devices that amplify quantum signals with minimal noise. These amplifiers are valuable in quantum communication and quantum measurement. These interactions can lead to interesting nonlinear optical effects in quantum systems which have applications in the generation of nonclassical light states and quantum nonlinear optical devices. In a nutshell, dispersive atom-field interactions are central to a wide range of quantum technologies which primarily motivated us to start this work. They enable precise control over atomic systems and have applications in quantum computing, sensing, optics, and the study of fundamental physics.  In this article, we consider a situation wherein a strong external coherent field (microwave or laser, depending on the type of the cavity QED experiment) resonant with a cavity mode, interacts non resonantly, i.e., dispersively, with a three-level atom passing through the cavity.

The study of nonclassical light continues to be an active and exciting field of research in quantum optics with the potentiality for significant technological advancements in the future. These states have numerous applications in different regimes. For example, quantum cryptography, quantum metrology, quantum computing, quantum imaging, and quantum sensing are many areas where nonclassical states have been applied with great success. In quantum cryptography \cite{cr}, nonclassical light sources such as single-photon sources \cite{yuan} are crucial for implementing quantum key distribution protocols \cite{keyd}. These protocols enable secure communication by allowing the exchange of cryptographic keys between two parties that cannot be intercepted without being detected. Squeezed states can be used to improve the precision of measurements beyond the standard quantum limit \cite{motion}, which can be in fields such as gravitational wave detection and atomic clocks. Nonclassical states can be used as qubits in quantum computing \cite{qc,qc1,qc2,qc3,qc4,qc5}, where they enable faster and more efficient processing of quantum information \cite{qi} as compared to classical bits.

In this paper, we propose an extended driven Jaynes-Cummings model where the cavity and external driving field are close to or on resonance with the atom. The external field, resonant with the cavity mode, interacts with an appropriately prepared atom as it passes through the cavity. The resulting interaction is conditional on the state of the atom and effectively converts the classical external field into a quantized cavity field with the same frequency. It can be seen that when the injected atom is prepared in a superposition of its bare atomic states, it becomes possible to generate various types of Schr\"{o}dinger cat states \cite{garry}. 

This paper is organized as follows. In Sect.~\ref{sec1}, we discuss the driven Jaynes-Cummings model in the regime where the
atom is detuned with both the quantized cavity field
and driving external classical field such that the atom-cavity field coupling is dispersive.
We study nonclassical properties of the cavity field state via a set of nonclassicality witnesses, namely zeros of $Q_f$ function, Mandel's $Q_M$ parameter, negativity of Wigner function, squeezing properties,  second-order correlation function ($g^2(0)<1$) in Sect.~\ref{sec2}.

	\section{State of interest}
	\label{sec1}
\begin{figure}[htb]
	
		\centering
		\includegraphics[width=0.4\textwidth]{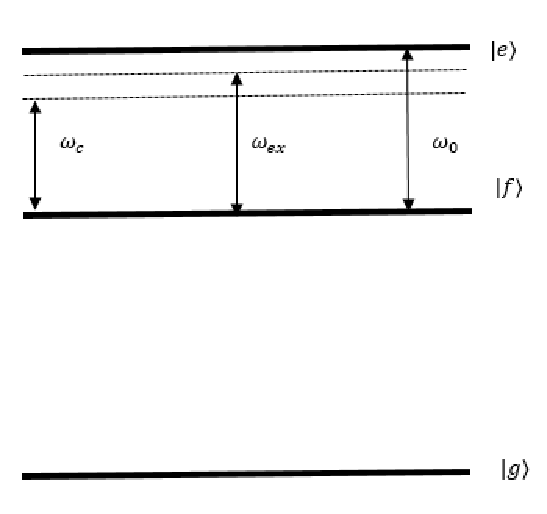}
	\caption{
		The system under consideration involves an atom interacting with both a cavity field and an external driving field. The energy-level configuration of the atom consists of three distinct levels denoted as $\ket{e}$, $\ket{f}$, and $\ket{g}$. The transition frequency between levels $\ket{e}$ and $\ket{f}$ is represented by $\omega_0$. In addition, there are two distinct frequencies involved: $\omega_c$ corresponds to the cavity field  while $\omega_{ex}$ represents the frequency of the external driving field. The cavity and external fields are positioned close to resonance but not exactly with the atomic transition frequency $\omega_0$. This arrangement allows for a dispersive interaction between the fields and the atom. It is assumed that the condition $\omega_c=\omega_{ex}$ holds true. Furthermore, the level $\ket{g}$ is significantly detuned from all the frequencies involved in the system.}
		\label{state}
	\end{figure}

	We consider an atom with three levels $\ket e$, $\ket f$, and $\ket g$ configured as in Fig.~\ref{state}. We assume that only dipole type transitions can occur consecutively: $\ket e\leftrightarrow\ket f\leftrightarrow\ket g$. $\omega_0$ is the atomic transition frequency between the levels $\ket e$ and $\ket f$ and it is near resonance with a single-mode cavity field of frequency $\omega_c$. We further assume that the transition $\ket f\leftrightarrow\ket g$ is far out of resonance with the specific cavity mode of interest or any other cavity mode. A strong, classical driving field of frequency $\omega_{ex}$  interacts directly with the atom while passing through the cavity.

The 	Hamiltonian for a system of a three-level atom (represented by Pauli matrices $\hat{\sigma}_i$) interacting with a quantized electromagnetic field mode (represented by the annihilation and creation operators ${a}$ and ${a}^\dagger$, respectively) is given by \cite{garry}
	\begin{align}
		\nonumber
		{H}&=\frac{1}{2} \hbar\omega_0\hat{\sigma}_3+\hbar\omega_c{a}^\dag a+\hbar g({a}\hat{\sigma}_++ {a}^\dag\hat{\sigma}_- )\\
		&+\hbar(Ee^{-i\omega_{ex}t}\hat{\sigma}_++E^*e^{i\omega_{ex}t}\hat{\sigma}_-)
	\end{align} 
Here $\frac{1}{2}\hbar\omega_0\hat{\sigma}_3$ is the energy of the atom in absence of any external fields where $\hat{\sigma}_3$ is the third Pauli matrix representing the difference in population between the two considered levels, $\hbar\omega_c{a}^\dag a$ is the energy of the electromagnetic field mode in absence of any atom, $a^\dagger$ and $a$ are the creation and annihilation operators for the field mode, $\hbar g({a}\hat{\sigma}_++ {a}^\dag\hat{\sigma}_- )$ is the interaction Hamiltonian between the atom and the cavity field, where $\hbar g$ is the coupling strength of the interaction and $\hat{\sigma}_+$ and $\hat{\sigma}_-$ are the raising and lowering operators for the atom, respectively,
$\hbar(Ee^{-i\omega_{ex}t}\hat{\sigma}_+ +E^*e^{-i\omega_{ex}t}\hat{\sigma}_-)$ is the external driving field that is applied to the atom, where $E$ is proportional to the coupling constant between the atom and the amplitude of the external classical field of frequency $\omega_{ex}$. This term can be used to drive transitions between the two levels of the atom. 

We are currently assuming that $\omega_0$, $\omega_c$ and $\omega_{ex}$ are distinct. To remove the time dependence in the Hamiltonian $H$, the system is rotating at the frequency $\omega_{ex}$. The modified Hamiltonian in the rotating frame is obtained as\\
\begin{align*}
	H_R&=i\hbar\dot{R}(t)R^\dag(t)+R(t)H(t)R^\dag (t)\\
	&=\frac{1}{2}\hbar \Delta\hat{\sigma}_3+\hbar(w_{c}-w_{ex})a^\dag a\\
	&+\hbar g\left(e^{-i\Delta t}\left({a}+\lambda\right)\hat{\sigma}_+
	+e^{i\Delta t}\left(a^\dag+\lambda^*\right)\hat{\sigma}_-\right)
\end{align*}
where $R(t) = \exp{-i\omega_{ex}t(\hat{\sigma}_3+a^\dag a)}$ is the rotation operator, $\lambda=\frac{E}{g}$, $\Delta = \omega_0-\omega_{ex}$ is the detuning between the atomic transition frequency and the external cavity field frequency. Assuming the resonance condition between the external field and the cavity field and introducing new auxiliary Bose operators ${b}={a}+\lambda$ and ${b}^\dag={a}^\dag+\lambda^*$ which satisfy $[{b}, {b}^\dag]=1$, the Hamiltonian can be written as
\begin{align}
	H_R&=\frac{1}{2}\hbar \Delta\hat{\sigma}_3+\hbar g\left(e^{-i\Delta t}{b}\hat{\sigma}_++e^{i\Delta t}{b^\dag}\hat{\sigma}_-\right)
\end{align}
which superficially looks like the interaction picture Hamiltonian of the usual detuned JCM. 

If the detuning between the atom and the fields is sufficiently large, one can use the standard techniques \cite{efh} to derive the effective atom-field interaction Hamiltonian as
\begin{align*}
	\int_0^{t}H_R(t')dt'	\int_{0}^{t'}H_R(t'')dt''
	\approx\frac{ih^2}{\Delta}[ b^\dag b \hat{\sigma}_3+\frac{1}{2}[b,b^\dag]\,(\hat{\sigma}_3+I)]t
\end{align*}
While performing the above integration, the second-order terms in $1/\Delta$ arriving from the exponential of $\Delta$ and an additional factor of $\Delta$ in the denominator are neglected. Here $\hat{\sigma}_3=\hat{\sigma}_+\hat{\sigma}_--\hat{\sigma}_-\hat{\sigma}_+=\ket e\bra e-\ket f\bra f$, $I=\ket e\bra e+\ket f\bra f$. Hence using $\hat{\sigma}_+\hat{\sigma}_-=\frac{1}{2}(\hat{\sigma}_3+I)$, the effective Hamiltonian arrives at
\begin{align}
	\nonumber
	H_\mathrm{eff}&=h\chi[\hat{\sigma}_+\hat{\sigma}_-+b^\dag b\hat{\sigma}_3]\\\nonumber
	&=h\chi[\hat{\sigma}_+\hat{\sigma}_-+D^\dag(\lambda)a^\dag aD(\lambda)\hat{\sigma}_3]\\
	&=h\chi[\hat{\sigma}_+\hat{\sigma}_-+(a^\dag a+\lambda a^\dag +\lambda^*a+|\lambda|^2)\hat{\sigma}_3]
\end{align}
where $\chi=\frac{g^2}{\Delta}$. In the limit $\lambda\rightarrow 0$ (in case of no external
driving field), the usual dispersive interaction
Hamiltonian $H_\mathrm{eff}=h\chi[\hat{\sigma}_+\hat{\sigma}_-+a^\dag a\hat{\sigma}_3]$ can be recovered. But with $\lambda\neq 0$, the interaction is no longer purely dispersive as it contains terms
of the form  that creates or destroys photons
in the cavity conditional on the state of the atom. 
If the atom is prepared in the far off-resonance state $\ket{g}$, the cavity field remains unaffected because there is no interaction between the atom and the field. But when the atom is prepared in either the excited state $\ket{e}$ or the intermediate state $\ket{f}$, and if the cavity field is initially in a vacuum state $\ket{0}$, the external classical driving field can generate a coherent state of the quantized field.
 
Assuming that the atom is initially prepared in state $\ket{f}$ and is injected through a cavity in the vacuum state $\ket{0}$, the atom-field system inside the cavity evolves as
 \begin{align}
 	\nonumber
 	\ket{\psi_f(t)}&=\exp[-iH_\mathrm{eff}t/\hbar]\ket{0}\ket{f}\\\nonumber
 	&=\exp[i\chi t(a^\dag a+\lambda a^\dag +\lambda^*a+|\lambda|^2)]\ket{0}\ket{f}\\\nonumber
 	&=\exp[i|\lambda|^2\sin(\chi t)]\ket{-\lambda(1-e^{i\chi t})}\ket f
 	 \end{align}
where $\ket{-\lambda(1-e^{i\chi t})}$ is a coherent state of the cavity field. Again if the atom is initially in the state $\ket{e}$, the state vector while the atom is inside the cavity becomes
\begin{align}
	\nonumber
	\ket{\psi_e(t)}&=\exp[-iH_\mathrm{eff}t/\hbar]\ket{0}\ket{e}]\\\nonumber
	&=\exp[-i\chi t-i\chi t(a^\dag a+\lambda a^\dag +\lambda^*a+|\lambda|^2)]\ket{0}\ket{e}\\\nonumber
	&=e^{-i\chi t}\exp[-i|\lambda|^2\sin(\chi t)]\ket{-\lambda(1-e^{-i\chi t})}\ket e
\end{align}
where $\ket{-\lambda(1-e^{- i\chi t})}$ is a  coherent state of the cavity field. If the atom is prepared in the general superposition
state $\sin\theta \ket{e}+e^{i\phi}\cos\theta\ket{f}$  and the cavity field in the vacuum
state, then at time $t\geq 0$, the following entangled state is obtained
$$\ket{\psi(t)}=\sin(\theta)\ket{\psi_e(t)}\ket{e}+e^{i\phi}\cos(\theta)\ket{\psi_f(t)}\ket{f}$$
For the simple most case $\theta={\pi}/{4}$ and $\phi=0$, the state vector becomes
\begin{eqnarray}
\ket{\psi(t)}=\frac{1}{\sqrt2}\left(\ket{\psi_e(t)}\ket{e}+\ket{\psi_f(t)}\ket{f}\right)
\end{eqnarray}
The state vector for the cavity field after the atom-field interaction is obtained by tracing out the atom part
from $\ket{\psi(t)}\bra{\psi(t)}$ as following:
$$Tr_\mathrm{atom}(\ket{\psi}\bra{\psi})=\ket{\psi_{\mathrm {field}}}\bra{\psi_{\mathrm {field}}}$$
$\ket{\psi_{\mathrm {field}}}$ is the state of interest in rest of the article.
\section{Generalised expectation}
The generalised expectation value with respect to the cavity field is obtained as
\begin{align}
	\nonumber
	\label{ex}
	&\bra{\psi_\mathrm{field}} a^{\dag p}a^q\ket{\psi_\mathrm{field}}=\frac{1}{2}\sum_{n=0}^{\infty}\frac{\lambda^n\lambda^{*(p+n-q)}}{(n-q)!}\\\nonumber
	&\times\bigg[\exp{-2|\lambda|^2
		(1-\cos( \chi t)}(1-e^{(i\chi t)})^{p+n-q}(1-e^{(-i\chi t)})^n\\\nonumber
	&+\exp(i\chi t)\exp{-2|\lambda|^2
		(1-e^{(i \chi t)})}(1-e^{(i\chi t)})^{p+2n-q}\\\nonumber
	&+\exp(-i\chi t)\exp{-2|\lambda|^2
		(1-e^{(-i \chi t)})}(1-e^{(-i\chi t)})^{p+2n-q}\\
	&+\exp{-2|\lambda|^2
		(1-\cos( \chi t)}(1-e^{(i\chi t)})^n(1-e^{(-i\chi t)})^{p+n-q}\bigg]
\end{align}
\section{Nonclassical Properties}
\label{sec2}
In this section, we derive some criteria to witness the nonclassicality of the considered quantum state.
\subsection{Photon number distribution}
\begin{figure}[htb]
	\centering
	\includegraphics[width=0.3\textwidth]{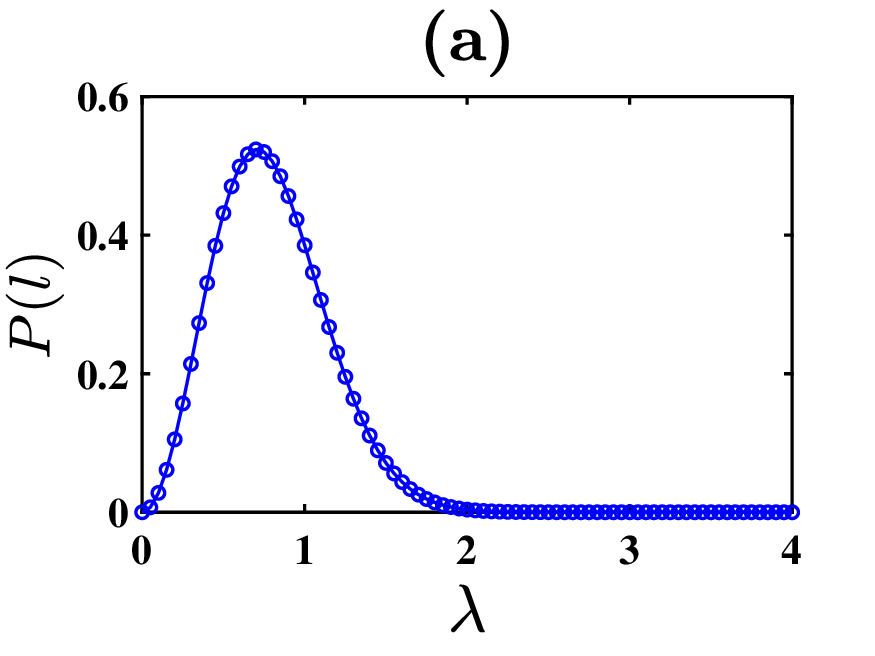}
	\includegraphics[width=0.3\textwidth]{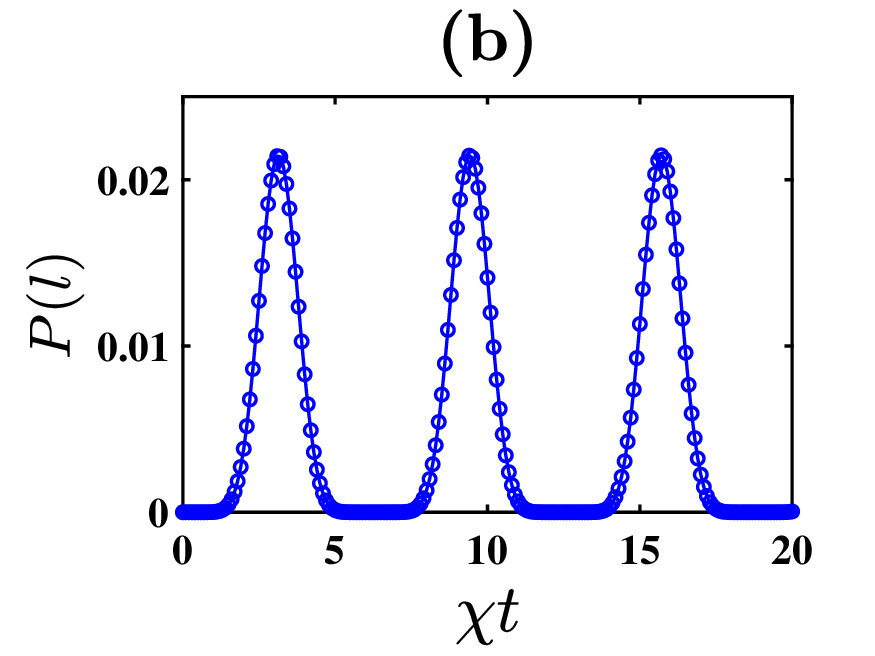}
	\includegraphics[width=0.3\textwidth]{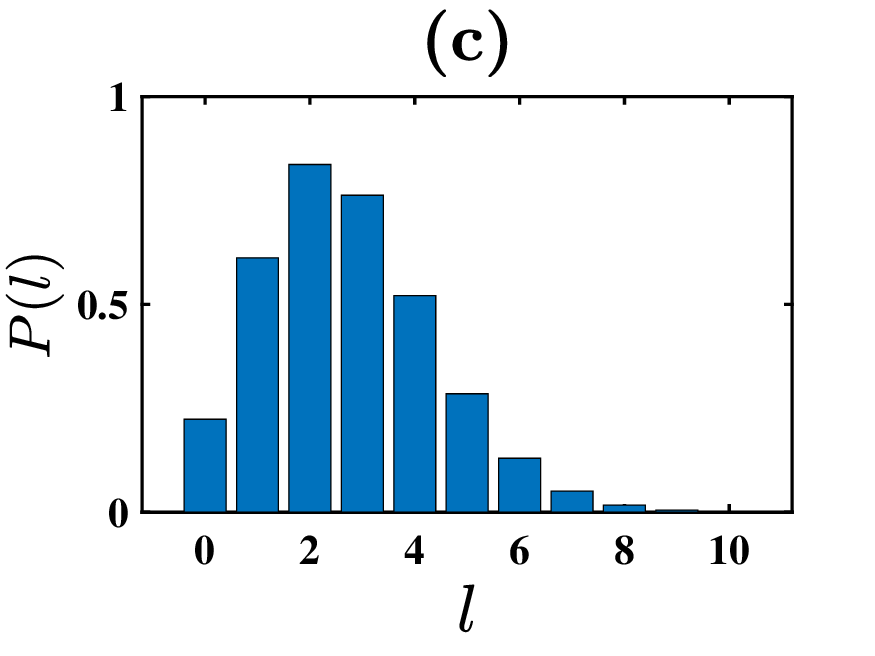}
	\caption{Variation of photon number distribution $P(l)$ as a function of (a) $\lambda$ with $\chi t=2,l=2$, (b) $\chi t$ with $\lambda=2,l=2$, (c) $l$ with $\lambda=1$ and $\chi t=4$.}
	\label{pn}
\end{figure}

	Photon number distribution is the probability distribution for finding $l$ photons in a given cavity field state and it can be obtained as the expectation of the density field in Fock state basis as
\begin{align*}
	P(l)&=\bra{l}\rho\ket{l}\\
	&=\frac{1}{2}\sum_{n=0}^{\infty}\frac{\lambda^n\lambda^{*m}}{\sqrt{n!m!}}\\
	&\times\bigg[\exp{-2|\lambda|^2
		(1-\cos( \chi t)}(1-e^{(-i\chi t)})^n(1-e^{(i\chi t)})^m\\
	&+\exp{-2|\lambda|^2
		(1-\cos( \chi t)}(1-e^{(i\chi t)})^n(1-e^{(-i\chi t)})^m\bigg]\\
	&\times\bra{l}\ket{n}\bra{m}\ket{l}
\end{align*}
For $m=n=l$, the photon number distribution for the cavity field is obtained as
\begin{align}
	\label{pn1}
	P(l)&=\frac{|\lambda|^n}{n!}\bigg[\exp{-|\lambda|^2
		(2-2\cos( \chi t)}(2-2\cos(\chi t))^{l}\bigg]
\end{align}
\eqref{pn1} provides the probability of observing $l$ number of photons in a given cavity field state. The simplified form of the photon number distribution is illustrated in Fig.~\ref{pn}. It is clear that $P(l)$ behaves in a wave-like manner as a function of $\lambda$ and $\chi t$. Initially $P(l)$ increases with $\lambda$ and then decreases. $P(l)$ attains a maximum value of 0.5 for $\lambda$, and 1.08 for $\chi t$. When plotting $P(l)$ for different photon numbers, the highest value of $P(l)$ is 0.83 for $l = 2$.

Here the shape of the photon number distribution provides insights into nonclassical behaviour of the considered quantum state as a consequence that nonclassical states may have unique features in their photon number distributions, such as multiple peaks or oscillatory behaviour. Experimental techniques and tools, such as photon-number-resolving detectors and homodyne or heterodyne detectors that measure photon number distributions with high precision are used in various experiments in quantum optics and quantum information science to verify the nonclassical nature of the quantum states.

\subsection{Mandel’s $Q_M$ parameter}
\begin{figure}[htb]
	\centering
		\includegraphics[width=0.3\textwidth]{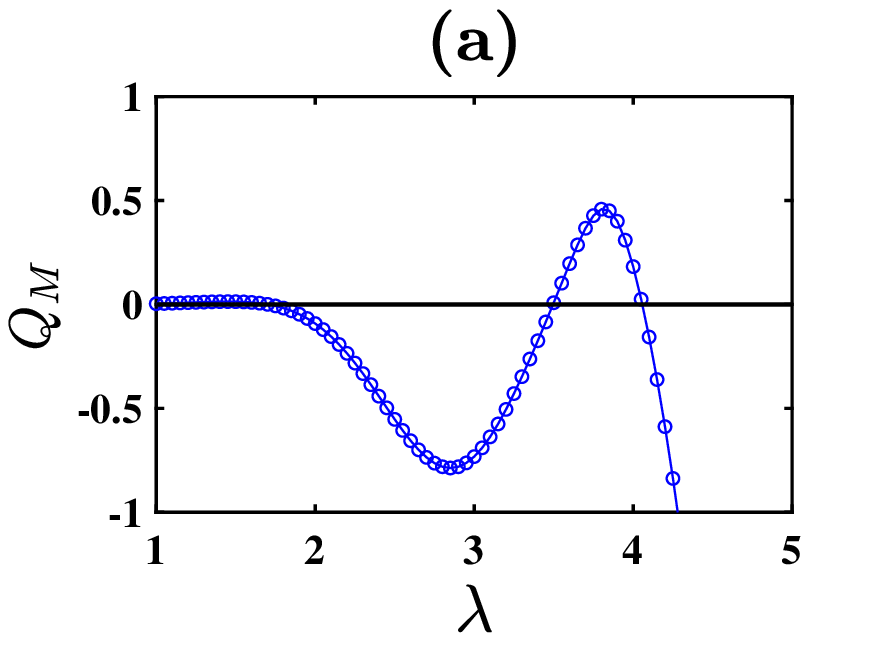}
	\includegraphics[width=0.3\textwidth]{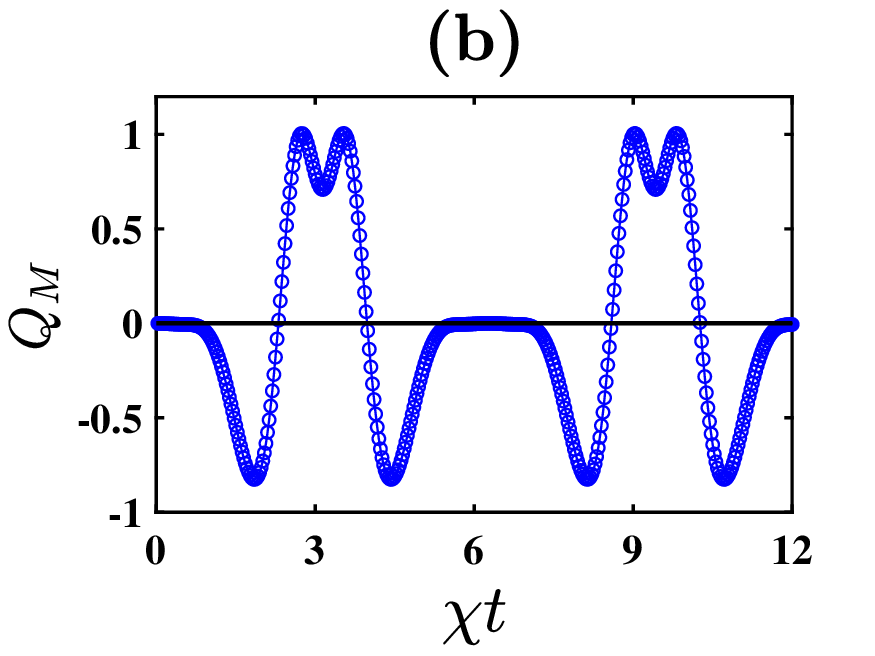}
	\caption{Variation of Mandel's $Q_M$ parameter as a function of (a) $\lambda$ with $\chi t=0.2$, (b) $\chi t$ with $\lambda=0.35$.}
	\label{Q}
\end{figure}

Next to determine the photon statistics of a single-mode radiation field, we consider the Mandel's $Q_M$
parameter defined by \cite{mandel}
\begin{equation}
	\label{eq14}
	Q_M=\frac{\langle a^{{\dagger}^2} a^2 \rangle}{\langle a^{\dagger}a\rangle} -\langle a^{\dagger}a\rangle
\end{equation}
The parameter $Q_M$ is used to quantify the deviation of the variance of the photon number distribution for a given state from the Poissonian distribution associated with a coherent state.
When $Q_M = 0$, it indicates a Poissonian distribution, whereas for $-1\leq Q_M<0~(Q_M>0)$, the field follows sub- (super-) Poissonian photon statistics. However, it is important to note that the negativity of $Q_M$ is not a necessary criterion to differentiate quantum states into classical and nonclassical regimes. It is merely a sufficient condition. There are cases where a state can exhibit nonclassical behaviour even if $Q_M$ is positive \cite{gs}. The expectation values can be found as following:
\begin{eqnarray*}
	\langle a^{{\dagger}^2} a^2 \rangle & = & \langle \psi_\mathrm{field}|a^{{\dagger}^2} a^2|\psi_\mathrm{field} \rangle\\
	\langle a^{\dagger}a\rangle & = & \langle \psi_\mathrm{field}|a^{\dagger} a|\psi_\mathrm{field} \rangle\\
\end{eqnarray*}
Substituting these in \eqref{eq14}, the expression for Mandel's $Q_M$ parameter for the given cavity field is obtained and plotted as a function of $\lambda$ and $\chi t$ in Fig.~\ref{Q}. Here we can see that $Q_M$ exhibits wave nature with varying amplitudes. The negativity of $Q_M$ ascertains nonclassical nature of the considered cavity field state. Also the nonclassicality decreases while $\lambda$ increases.  The state portrays sub-Poissonian light $(Q_M<0)$ for $1.8<\lambda<3.5$ and for specific values of $\chi t$ ranging from 0.70 to 11.7. Mandel's $Q_M$ parameter can be used in a variety of applications, including the characterization of nonclassical states of light, the measurement of photon number correlations in quantum states, and the implementation of quantum communication protocols. It is a useful tool for quantifying the degree of photon number correlations in a quantum state, and has played an important role in the development of quantum optics and quantum information science.
\subsection{Squeezing properties}
\begin{figure}[htb]
	\centering
	\includegraphics[width=0.3\textwidth]{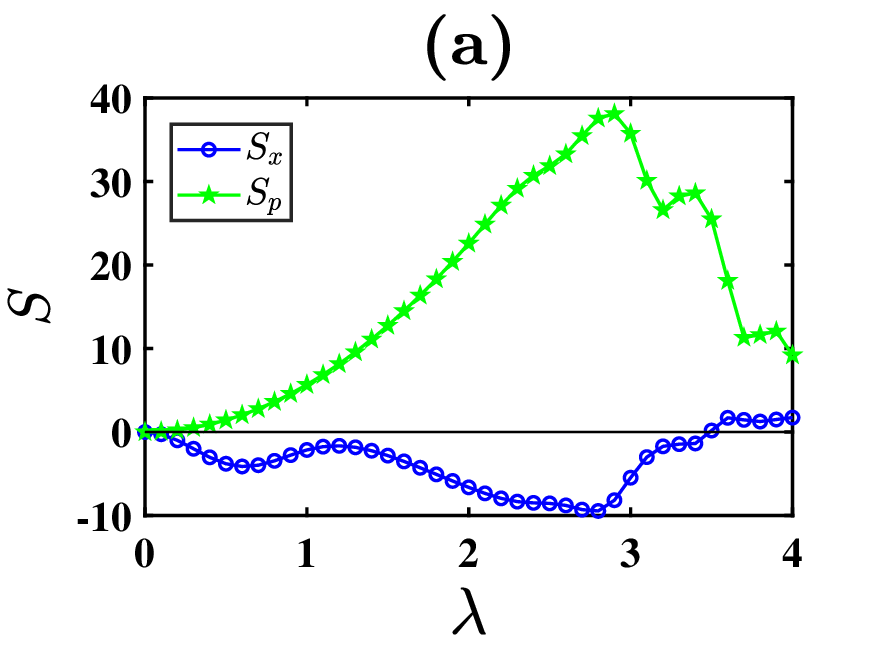}
	\includegraphics[width=0.3\textwidth]{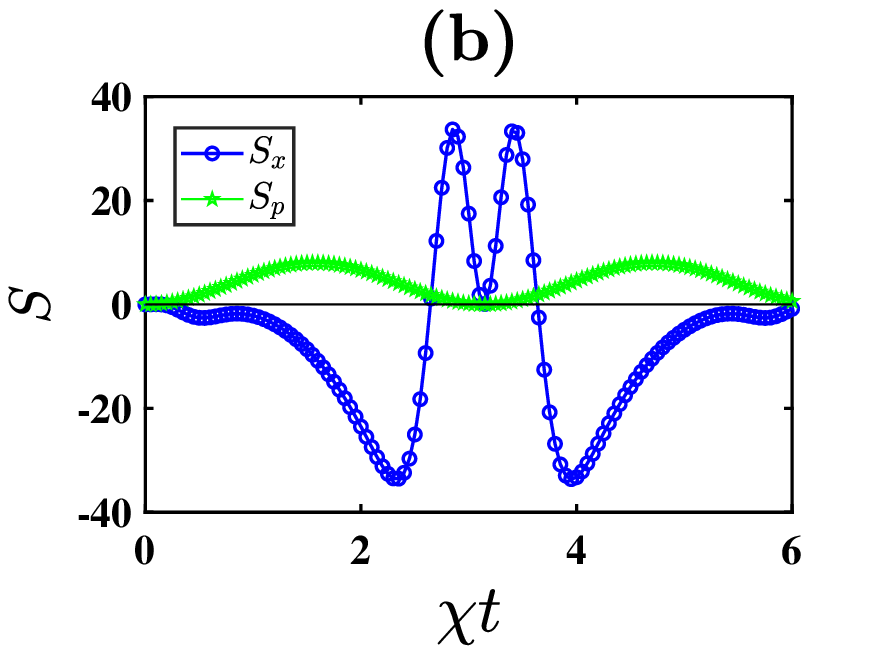}
	\caption{Variation of squeezing parameters $S=(S_x,S_p)$ as a function of (a) $\lambda$ with $\chi t=1$, (b) $\chi t$ with $\lambda=1$. }
	\label{S}
\end{figure}

To analyze the quantum fluctuations of the field quadratures, we consider two Hermitian operators which are combinations of photon creation and annihilation operators as following:

$$\hat{x}=\frac{{a}+{a}^{\dagger}}{2},\,\,\,\,\,\,\hat{p}=\frac{{a}-{a}^{\dagger}}{2i}$$\\
with the commutation relation $\left[\hat{x},\hat{p}\right] = i/2.$ They obey the Heisenberg uncertainty principle of the form
$\langle(\Delta\hat{x})^2\rangle\langle(\Delta\hat{p})^2\rangle\geq1/16$, and thus the quadrature squeezing occurs whenever $\langle(\Delta\hat{x})^2\rangle<\frac{1}{4}$ or $\langle(\Delta\hat{p})^2\rangle<\frac{1}{4}$. It is convenient to introduce the squeezing parameters as \cite{nk}
\begin{align*}
	S_x&=2\langle a^\dag a\rangle+\langle a^2\rangle+\langle a^{\dag 2}\rangle-{\langle a\rangle}^2-{\langle a^\dag\rangle}^2-2\langle a\rangle \langle a^\dag\rangle\\
	S_p&=2\langle a^\dag a\rangle-\langle a^2\rangle-\langle a^{\dag 2}\rangle+{\langle a\rangle}^2+{\langle a^\dag\rangle}^2-2\langle a\rangle \langle a^\dag\rangle\\
\end{align*}
 Squeezing occurs in the $x$ or $p$ quadrature when $-1 < S_x < 0$ or $-1 < S_p < 0$, respectively. The negativity of squeezing serves as a sufficient condition rather than a necessary one. The expectations can be computed using \eqref{ex}. We have plotted the squeezing values $S=(S_x,S_p)$ as a function of $\lambda$ and $\chi t$ in Fig.~\ref{S}. Here $S_x$ becomes negative as a function of $\lambda$ as well as $\chi t$, indicating the nonclassical nature of the given cavity field state.
Squeezed states of light has the potential to enhance the capabilities of quantum devices. They are crucial in the generation of entangled photon pairs, which are essential for various quantum technologies, including quantum teleportation and quantum cryptography. This light is proposed to improve the sensitivity of gravitational wave detectors in today's most referenced LIGO (Laser Interferometer Gravitational-Wave Observatory) experiments. It can enhance the precision of interferometric measurements of space-time distortions caused by gravitational waves.

\subsection{Wigner distribution}

The nonclassical characteristic of a quantum state can be detected by examining its phase-space distribution, namely the Wigner function. This function establishes a connection between the density operator and a distribution in phase space. It contains comprehensive information about the state of a particular physical system \cite{Wgnr}. This distribution is often referred to as a quasiprobability distribution because it has the ability to take on negative values. These negative values serve as indicators of the system's departure from classical behavior.

\begin{figure}[htb]
	\centering
		\includegraphics[width=0.3\textwidth]{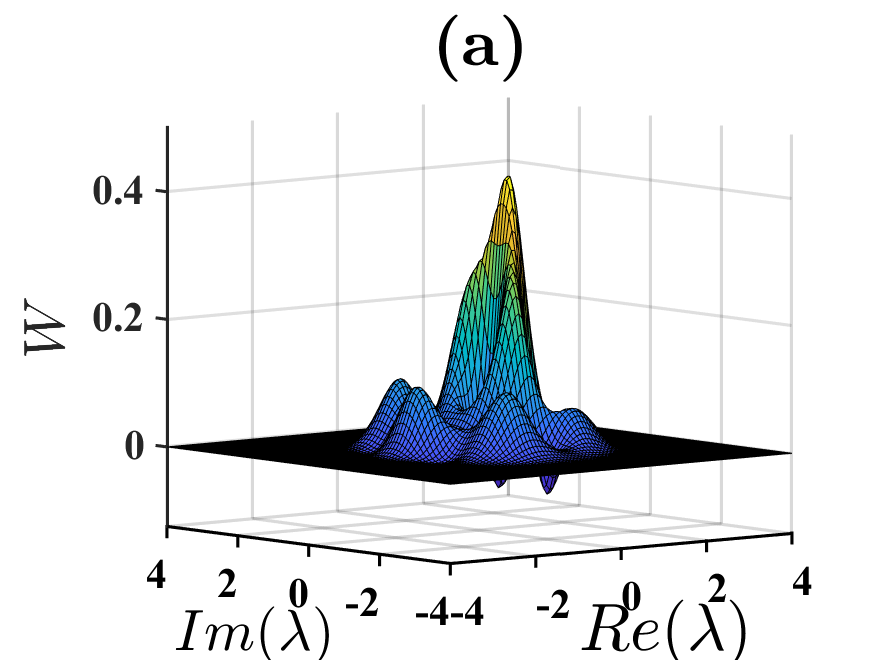}
	\includegraphics[width=0.3\textwidth]{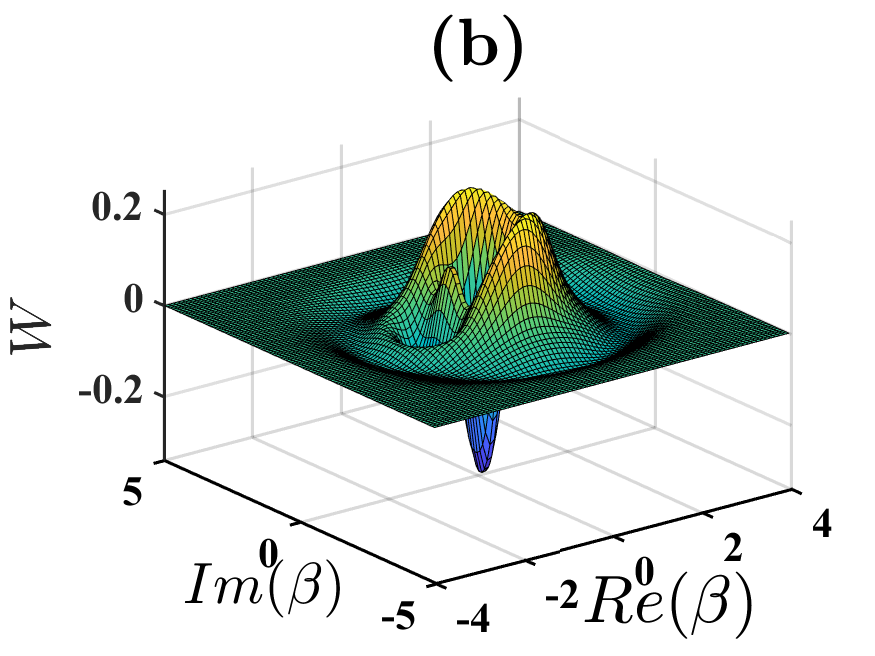}
		\includegraphics[width=0.3\textwidth]{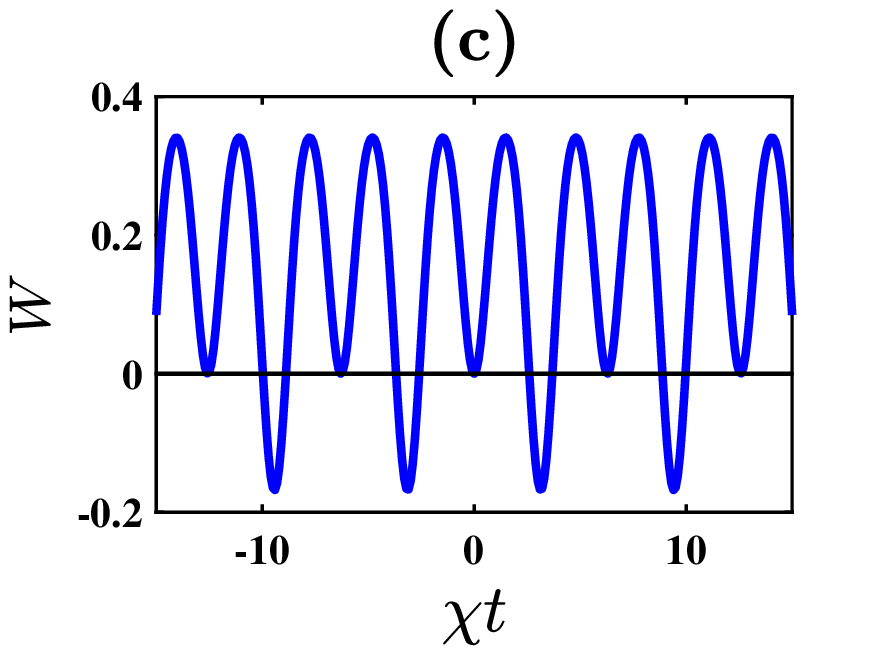}
	\caption{Variation of Wigner function $W$ with respect to (a) $\lambda$ with $\chi t=2$ and $\beta=1$, (b) $\beta$ with $\chi t=1$ and $\lambda=1$, (c) $\chi t$ with $\beta=\lambda=1$. }
	\label{w}
\end{figure}

Given a quantum state $\hat{\rho}$, the Wigner function of the system is defined using the coherent state basis as follows \cite{motion,pathak,chatterjee}
\begin{eqnarray*}
	W(\beta, \beta^*) = \frac{2}{\pi^2}e^{2|\beta|^2} \int d^2\gamma{\langle-\gamma|\hat{\rho}|\gamma\rangle e^{-2(\beta^*\gamma-\beta\gamma^*)}},
\end{eqnarray*}
where $|\gamma\rangle=\exp(-|\gamma|^2/2+\gamma \hat{a}^\dag)|0\rangle$ is a coherent state. By using the relation \cite{abramowitz72}
\begin{eqnarray*}
	\sum_{n=k}^\infty n_{C_k}\,y^{n-k} = (1-y)^{-k-1},
\end{eqnarray*}
the Wigner function can be expressed in a series form as follows \cite{moyacessa93},
\begin{eqnarray}
	\label{eqn8}
	W(\beta, \beta^*) = \frac{2}{\pi} \sum_{k=0}^\infty (-1)^k \langle \beta,k|\hat{\rho}|\beta,k\rangle ,
\end{eqnarray}
where $|\beta,k\rangle$ is the standard displaced number state. The presence of negativity in the Wigner function indicates that the associated state is nonclassical \cite{wangz}. But observing positive values for the entire Wigner function cannot lead to the conclusion that the state is classical. For example, squeezed state owning a Gaussian Wigner function that is positive everywhere, is a widely recognized nonclassical state. In the case of a nonclassical state, the negativity of the Wigner function is a necessary condition. Therefore, a state exhibiting a negative region in its phase-space distribution is inherently nonclassical. The displaced number state $|\beta, k\rangle$ can be alternatively represented in the number state basis as follows,

\begin{eqnarray}\nonumber
	\label{eqn9}
	|\beta, k\rangle & = &
	D(\beta)| k \rangle\\\nonumber
	& = & e^{-\frac{|\beta|^2}{2}}\sum_{l=0}^{k} \frac{ \beta^{*{l}}}{l!} \sqrt{\frac{k!}{(k-l)!}}\\
	& & \times\sum_{p=0}^{\infty} \frac{ \beta^p}{p!}\sqrt{\frac{(k-l+p)!}{(k-l)!}}|k-l+p \rangle
\end{eqnarray}
Thus
\begin{eqnarray}\nonumber
	& & \langle \beta,k|n \rangle=
	e^{-\frac{|\beta|^2}{2}}\sum_{l=0}^{k} \frac{(-\beta)^{l}}{l!} \sqrt{\frac{k!}{(k-l)!}}\\\nonumber
	& & \times \sum_{p=0}^{\infty}\frac{ \beta^{*{p}}}{p!}  \sqrt{\frac{(k-l+p)!}{(k-l)!}}\langle k-l+p|n\rangle\\\nonumber
	& = & e^{-\frac{|\beta|^2}{2}}\sum_{l=0}^{k} \frac{(-\beta)^{l}}{l!} \sqrt{\frac{k!}{n!}} \frac{(\beta^*)^{n-k+l} n!}{(k-l)!(n-k+l)!}\\\nonumber
	& = & e^{-\frac{|\beta|^2}{2}} \sqrt{\frac{k!}{n!}} (\beta^*)^{n-k}\sum_{l=0}^{k}  \frac{(-|\beta|^2)^{l} n!}{(k-l)!l!(n-k+l)!}\\
	& = & e^{-\frac{|\beta|^2}{2}} \sqrt{\frac{k!}{n!}} (\beta^*)^{n-k}L_{k}^{(n-k)}\left(|\beta|^2 \right),
\label{eq5}
\end{eqnarray}
where $L_{l}^{(k)}(x)=\sum_{n=0}^{l}  \frac{(-x)^{n} (l+k)!}{(l-n)! n!(l+n)!}$ is the associated Laguerre polynomial \cite{agg}. Now
\begin{align}
	\label{wf}
&\nonumber\langle \beta,k|\hat{\rho}|\beta,k\rangle = \frac{1}{2}\sum_{m,n=0}^{\infty}\frac{\lambda^n\lambda^{*m}}{\sqrt{m!n!}}\\\nonumber
&\times\bigg[\exp{-2|\lambda|^2
	(1-\cos( \chi t)}(1-e^{(-i\chi t)})^n(1-e^{(i\chi t)})^m \\\nonumber
&+\exp{-2|\lambda|^2
	(-\cos( \chi t)}(1-e^{(i\chi t)})^n(1-e^{(-i\chi t)})^m\bigg]\\&\times \langle \beta,k\ket{n}\bra{m}\beta,k\rangle&
\end{align}
Substituting \eqref{wf} into \eqref{eqn8}, we get the final expression of the Wigner function. The surface plot of the Wigner function $W$ with respect to $\lambda$ and $\beta$ is given in Fig.~\ref{w}. It is evident that $W$ exhibits negative regions as well as peaks, indicating the nonclassical and non-Gaussian characteristics of the cavity field. In addition, the Wigner function demonstrates wavy nature and goes negative with respect to the parameter $\chi t$. It is important to note that the identification of negative values within the Wigner function is a clear indicator of non-Gaussian and nonclassical characters.  These type of states hold great significance in the field of quantum information applications, mainly due to their resistance to effective simulation by classical computers \cite{VV,AM}.

\subsection{$Q_f$ Function}
\begin{figure}[h]
	\centering
	\includegraphics[width=0.3\textwidth]{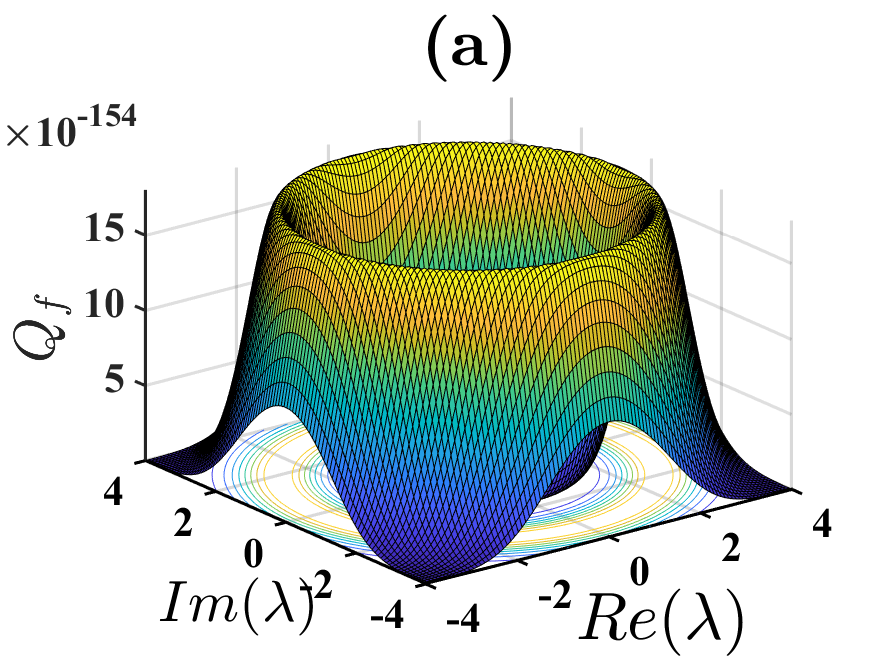}
	\includegraphics[width=0.3\textwidth]{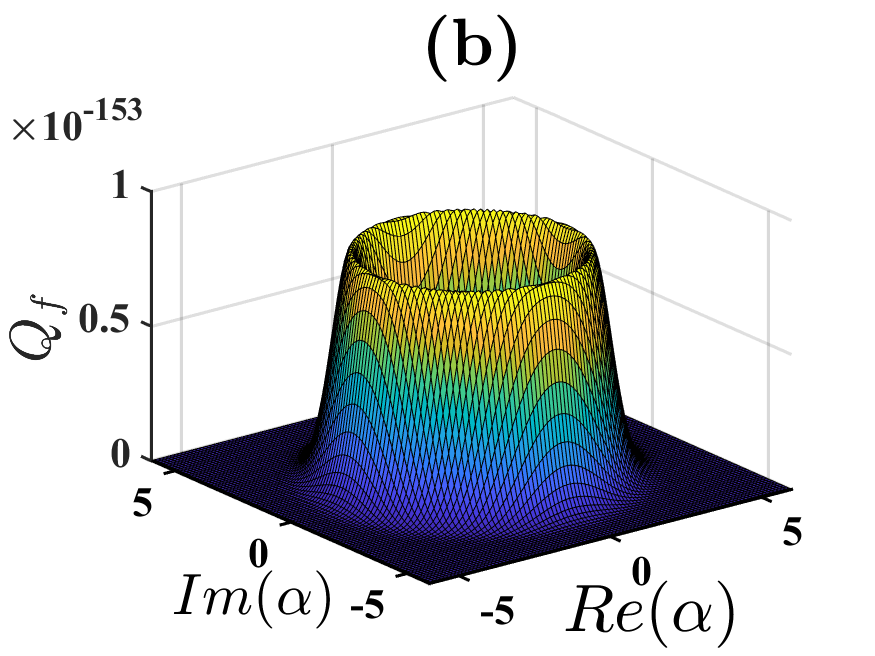}
	\includegraphics[width=0.3\textwidth]{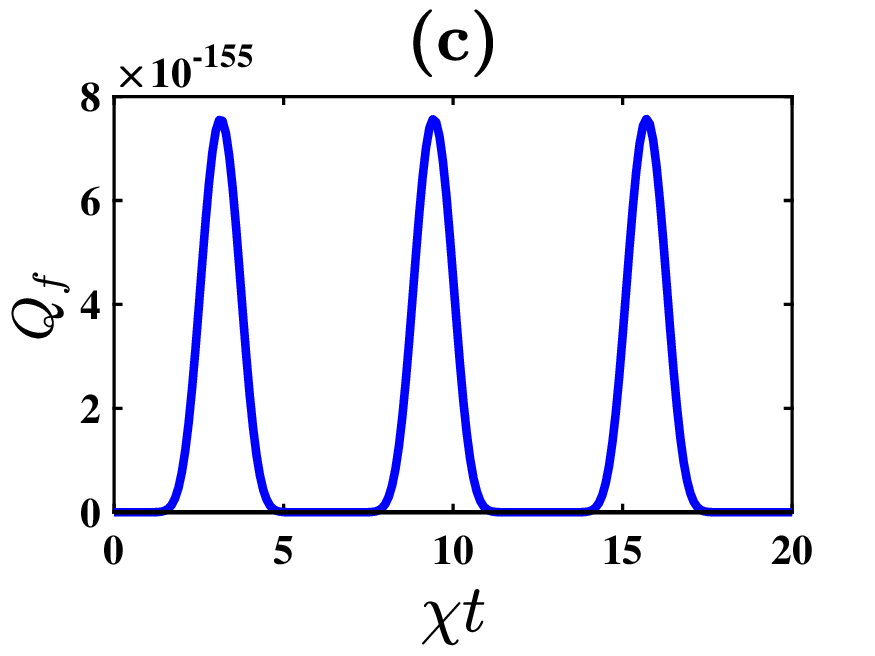}
	\caption{Variation of $Q_f$ function with respect to (a) $\lambda$ with $\chi t=1$ and $\alpha=1$, (b) $\alpha$ with $\chi t=1$ and $\lambda=1$, (c) $\chi t$ with $\alpha=1$ and $\lambda=1$.}
	\label{qfunction}
\end{figure}

The uncertainty principle imposes limitation on directly describing a quantum mechanical system in phase-space. Consequently, quasiprobability distributions have been introduced, which are highly valuable in quantum mechanics as they establish a correspondence between the quantum and classical realms. They also facilitate the calculation of quantum mechanical averages in a manner similar to classical phase-space averages \cite{qf}. One notable quasiprobability distribution is the $Q_f$ function, where the presence of zeros serves as an indicator of nonclassical behavior \cite{qf1}. The $Q_f$ function can be computed using the following expression
\begin{equation}
	\label{qf}
	Q_f=\bra \alpha \hat{\rho} \ket \alpha
	\end{equation}
where $\ket{\alpha}$ is the usual coherent state.
Substituting $\ket{\alpha}=e^{-\frac{|\alpha|^2}{2}}\sum_{p=0}^{\infty}\frac{\alpha^{p}}{\sqrt{p!}}\ket{p}$ in (\ref{qf}), we obtain 
\begin{align*}	
	Q_f&=\frac{1}{2}\sum_{m,n=0}^{\infty}\frac{\lambda^n\lambda^{*m}}{\sqrt{m!n!}}\\
	&\times\bigg[\exp{-|\lambda|^2
		(2-2\cos( \chi t)}(1-e^{(-i\chi t)})^n(1-e^{(i\chi t)})^m \\
	&+\exp{-|\lambda|^2
		(2-2\cos( \chi t)}(1-e^{(i\chi t)})^n(1-e^{(-i\chi t)})^m\bigg]\\
	&\times\bra \alpha\ket{n}\bra{m}\ket{\alpha}
\end{align*}
where
\begin{align*}
\bra	\alpha\ket{n}\bra{m}\ket{\alpha}
	&=\sum_{p=0}^{\infty}\frac{\alpha^{*p}}{\sqrt{p!}}e^{-\frac{|\alpha|^2}{2}}\sum_{q=0}^{\infty}\frac{\alpha^{q}}{\sqrt{q!}}e^{-\frac{|\alpha|^2}{2}}\bra{p}\ket{n}\bra{m}\ket{q}\\
	&=e^{-{|\alpha|^2}}\frac{\alpha^{*n}}{\sqrt{n!}}\frac{\alpha^{m}}{\sqrt{m!}}
\end{align*}
$Q_f$ is a complex-valued function that can be plotted in the complex plane, and it provides a complete description of the quantum state of a system. The surface plot of $Q_f$ as a function of $\lambda$ and $\alpha$ is shown in Fig.~\ref{qfunction}. It is clearly seen that $Q_f$ almost overlaps with zero for certain values of the parameters, which indicates the nonclassical behaviour of the cavity field state. This function has a number of important applications in quantum optics, quantum information processing, and quantum computing. For example, it can be used to calculate the probability of detecting a photon in a particular mode of a quantum field, or to describe the entanglement of two or more quantum systems.

\subsection{Second-order correlation}
The second-order correlation function $g^2(0)$ at zero time
delay for a single-mode radiation field is defined as \cite{g2}\\
\begin{equation}
	g^2(0)=\frac{\bra{\psi_\mathrm {field}}a^{\dag 2}a^2\ket{\psi_\mathrm {field}}}{\bra{\psi_\mathrm {field}}a^{\dag}a\ket{\psi_\mathrm {field}}^2}
\end{equation}
If for a field state $g^2(0)>1$,
the field is said to be bunched with super-Poissonian field statistics, whereas if
$g^2(0)<1$, the field is said to be antibunched with sub-Poissonian field statistics. 
\begin{figure}[htb]
	\centering
	\includegraphics[width=0.3\textwidth]{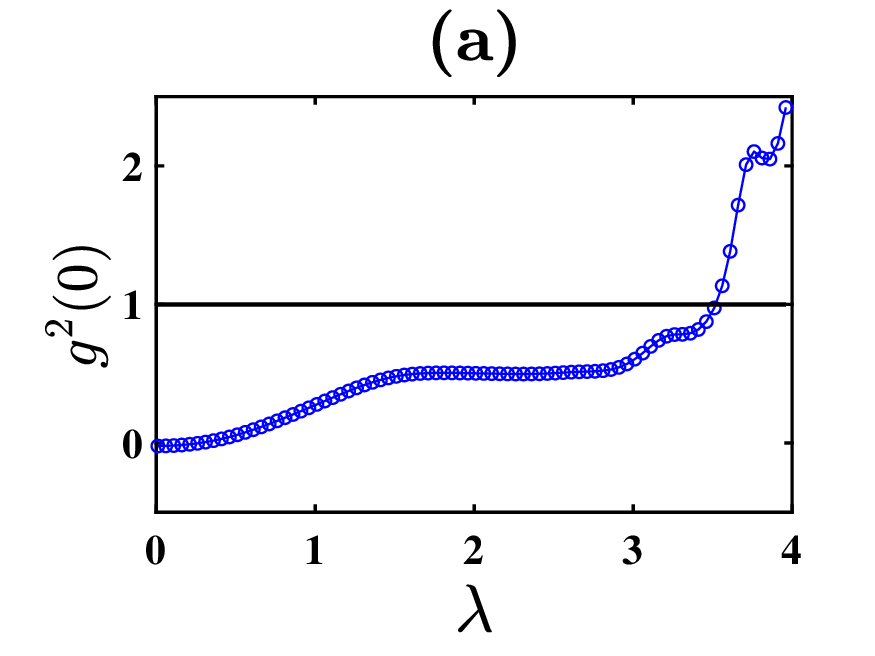}
	\includegraphics[width=0.3\textwidth]{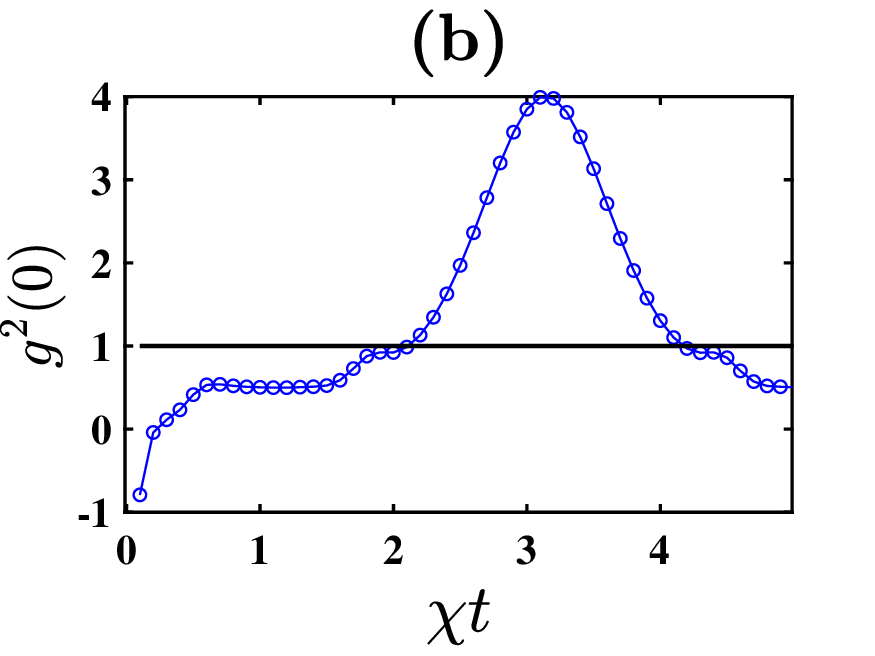}
	\caption{Variation of second-order correlation $g^2(0)$ as a function of (a) $\lambda$ with $\chi t=1$, (b) $\chi t$ with $\lambda=2$.}
	\label{g2}
\end{figure}
We have seen from Fig.~\ref{g2} that $g^2(0)$ is less then one for some particular values of $\lambda$ as well as $\chi t$ which imparts that the cavity field state is antibunched.
\subsection{Lower order antibunching}
\begin{figure}[htb]
	\centering
	\includegraphics[width=0.3\textwidth]{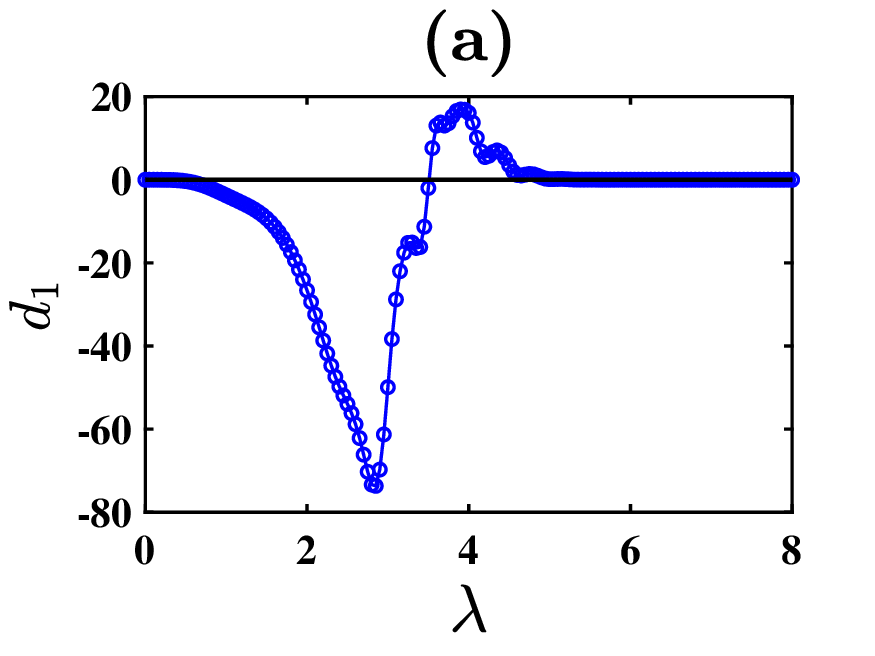}
	\includegraphics[width=0.3\textwidth]{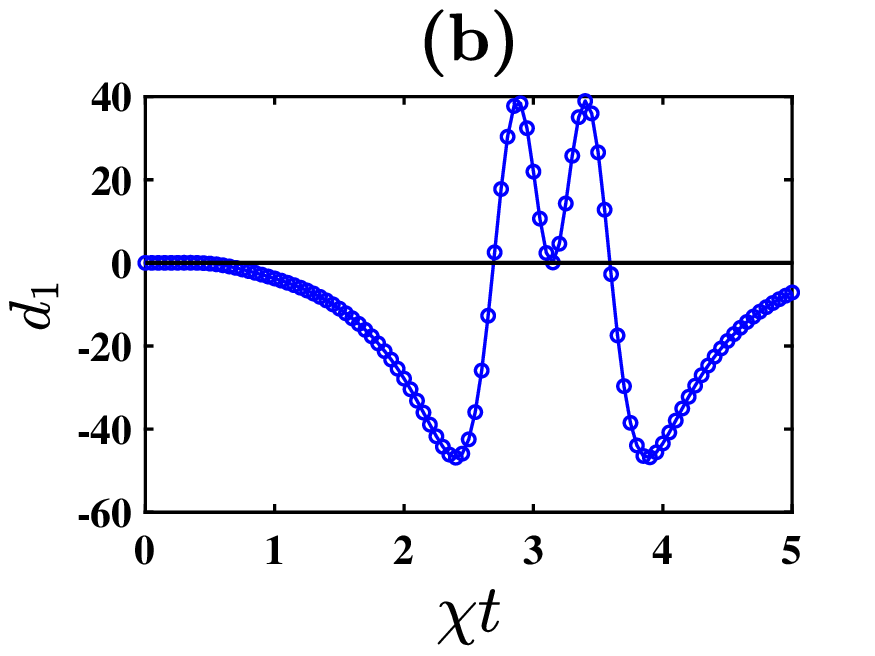}
	\caption{The variation of the lower order antibunching $d_1$  as a function of (a) $\lambda$ with $\chi t=1$, (b) $\chi t$ with $\lambda=2$.}
	\label{Ab}
\end{figure}

The theory of majorization by \cite{lee} and \cite{path} gives us the expression of antibunching as
\begin{equation*}
	d_{1}=\langle a^{{\dagger}^2} a^2 \rangle -{\langle a^{\dagger}a\rangle}^2
\end{equation*}
For a quantum state to be nonclassical, $d_{1}$ should be negative. The negativity of $d_{1}$ implies that the probability of independent photons is more than the clustered photons. Here $d_{1}$ can simply be calculated by substituting the expectations $\langle a^{{\dagger}^2}a^2\rangle$ and $\langle a^{\dagger}a\rangle$ from \eqref{ex}. Fig.~\ref{Ab} clearly shows that $d_1$ becomes negative with respect to $\lambda$ and $\chi t$, indicating that the cavity field state is antibunched. We have observed that the cavity QED field exhibits antibunching behaviour for a short period of time that means the system presents signature
	of temporary antibunching. In many experiments, it is important to stabilize the antibunching effect over time to ensure that
	the nonclassical properties of the light source are maintained.  Temporary antibunching can also be used to generate entangled photon pairs in a
	random sequence which is necessary for the success of teleportation protocol. Thus temporary antibunching present for a
	limited time period is expected to play an important role in the
	study of cavity QED system.
\section{Conclusion}
In this paper, we have investigated the generation of a state by the interaction of an atom with both a quantized cavity field and an external classical driving field. In this study, we have considered an atom with three levels $\ket e$, $\ket f$ and $\ket g$ passes through a cavity and interacts dispersively with the cavity field mode. At the same time it interacts with an external classical field tuned into resonant with the cavity field. We have further assumed that the transition $\ket f\leftrightarrow\ket g$ is significantly far from being in resonance and the transition frequency for $\ket e\leftrightarrow\ket f$ is in close proximity to resonance with the frequency of a single-mode cavity field. A coherent state of quantized field is generated, when the atom prepared in either states $\ket e$ or $\ket f$, and the cavity field is initially in vacuum state $\ket 0.$ We have considered that the atom is initially in the superposition of these two states and the cavity is in vacuum state. After the atom-cavity interaction, the cavity field state is obtained by tracing out the atom part from the generalized state vector. Different statistical properties like photon number distribution, Wigner function, Mandel's $Q_M$ parameter, squeezing properties $S_x$ and $S_p$, $Q_f$ function, second-order correlation $g^2(0)$ etc. are investigated. It is observed that Wigner function $W$, Mandel's $Q_M$, squeezing ($S_x$, $S_p$) are negative for some specific parametric values that admits the nonclassical nature of the cavity field state. Also the zeroes of $Q_f$ function suggests that the behaviour of the cavity state is nonclassical. The second-order correlation function $g^2(0)$ with respect to $\lambda$ is less than 1 that also imparts the nonclassical behaviour of the cavity field.

In particular, we have investigated photon number distribution, quadrature squeezing, Mandel's $Q_f$ etc. and provided analytical expressions for these in dependence on different state parameters. Furthermore, the Wigner function of the cavity-mode
is calculated and we could infer the state's nonclassical
character from its negativities. For the preparation of states
with different nonclassical features optimal parameter regions
are identified. Summing up, the presented approach can
easily be implemented in current cavity QED experiments and
provides a versatile method for the engineering of nonclassical
states of cavity fields. Dispersive interactions can be used to create quantum networks where information is stored and processed using atoms trapped in optical cavities. Future research may focus on scaling up these networks and improving their efficiency for secure quantum communication.

\begin{center}
\textbf{ACKNOWLEDGEMENT}
\end{center}
Naveen Kumar acknowledges the financial support from the Council of Scientific and Industrial Research, Govt. of India (Grant no. 09/1256(0004)/2019-EMR-I).

\begin{center}
\textbf{Data Availability Statement}
\end{center}
Data generated or
analyzed during this study are provided in full within the article.

\end{document}